___

## A Formal Calculus for International Relations Computation and Evaluation

Mohd Anuar Mat Isa[1], Ramlan Mahmod[1], Nur Izura Udzir[1], Jamalul-lail Ab Manan[1], Audun Jøsang[2], Ali Dehghan Tanha[3]

1. Faculty of Computer Science and Information Technology, Universiti Putra Malaysia, Malaysia.
2. Department of Informatics, University of Oslo, Oslo, Norway.
3. School of Computing, Science & Engineering, University of Salford, Salford, United Kingdom.
Corresponding Author email: anuarls@hotmail.com

**ABSTRACT:** This publication presents a relation computation or calculus for international relations using a mathematical modeling. It examined trust for international relations and its calculus, which related to Bayesian inference, Dempster-Shafer theory and subjective logic. Based on an observation in the literature, we found no literature discussing the calculus method for the international relations. To bridge this research gap, we propose a relation algebra method for international relations computation. The proposed method will allow a relation computation which is previously subjective and incomputable. We also present three international relations as case studies to demonstrate the proposed method is a real-world scenario. The method will deliver the relation computation for the international relations that to support decision makers in a government such as foreign ministry, defense ministry, presidential or prime minister office. The Department of Defense (DoD) may use our method to determine a nation that can be identified as a friendly, neutral or hostile nation.

**Introduction**

This publication describes an extension of our previous works related to trust issues in international nation relations. The trust issues become prominent due to a complexity of international relations in digital world. To explore the trust issues, we have investigated existing literatures that discuss trust (e.g. definition, formalism and theory) and international relations. From an observation in the literatures, we found none of the existing literatures discussed a calculus model for the international relations. Many literatures in politics and social sciences discussed the trust and international relations in philosophical forms which are subject to ambiguity of natural languages (J.Wheeler, 2012; Jelenc and Trček, 2014; Mcknight and Chervany, 2001). A few literatures in computer science and applied mathematics discuss trust as mathematical models (Hallerstede et al., 2014; Jøsang, 1997). However, the trust mathematical models in existing literatures are too general and it is not sufficient to be used in modeling the trust issues in international nation relations. Probabilistic theories in trust modeling work such as Bayesian, Dempster-Shafer, Josang's subjective logic (Abdel-hafez, 2013; Dempster, 1967; Jøsang, 2013) that do not directly address the trust issues for international relations. A brief discussion regarding the previous literatures and its related theories are discussed in the Literature Review section. In this work, we model the international relations between nations using a calculus model, which we call a relation algebra. The proposed method will allow a relation computation, which is previously subjective and incomputable.

In our previous works (Mohd Anuar Mat Isa et al., 2012a, 2012b), we have mentioned the need for a "*trust model*" in Common Criteria (CC). T*rust* is an important element to ensure Common Criteria's participant nations are able to recognize and consume Common Criteria's products. Our previous work motivation is influenced by Kallberg's suggestion that *"the long-term survival of CC requires abandoning the global approach and instead use established groupings of trust"* (Kallberg, 2012). We also mentioned about game theory as a *strategic decision-making engine* to evaluate the trust model. However, our preceding publications did not describe the implementation of relation algebra for international nation relations model. This publication covers the *relation algebra for international nation relations model*. The aim of this work is to propose the relation algebra for relation computations and trust modeling in the international nation relations. The proposed relation algebra is then evaluated using case studies. The proposed relation algebra will provide mathematical evaluations of international relations between nations.

The outline of this paper is a follows: Our research motivations and objectives are presented in Motivation section, while Literature Review section examinations literature related to trust and international relations. The research methodology for the relation algebra is described in Methodology section. We then explain the relation algebra method, definition and notation in Trust Algebra section. In Case Study section, we test the relation algebra using public data (e.g. internet and research literature) as case studies for the international nation relations. We provide results and discussions of the case studies in Result and Discussion sections, while our research contributions and future works in Contribution and Future Work section. Finally, Conclusion section summarizes this research works.

**Related Work**

In this section, trust definitions and its modeling works will be discussed in the sub-sections.

*Trust Relation Definition*

The keyword "*trust*" and "*relation*" provide many definitions in various disciplines. Searching the keywords in a search engine reveals varieties of trust definitions. The definitions can be linked to economy and political science studies in the earliest stage of trust definition. Later the definitions have evolved due to its expansion in computer science studies. A trust relation definition in the social science can be categorized as an expectation about other motives in a specific situation or state (Mcknight and Chervany, 1996). This definition is rooted in relationships between two entities (trustor and trustee); a willingness to be vulnerable or taking a risk (Schoorman et al., 2007); a belief or faith about honesty (Mcknight and Chervany, 2001, 1996), benevolence of another entities in relationships (Porta and Scazzieri, 1997); and an experience and knowledge that are cultivated over time between entities (Jiang et al., 2015; Kydd, 2005).

In this paragraph, we focus on trust relation definitions that are related to international relations which are the main theme in this publication. Trust can be defined as a clustering of perception (White, 1992); an optimism about trustworthiness of others in mutual beneficial relations (Hardin, 1992); an expression of people in evaluating others in any given society and its evaluation will be affected based on changes in the external world (Newton, 2001); and a trust level can be mapped to the amount of risk one is willing to take in relations (Schoorman et al., 2007). A variety of trust quotations from the international relations literatures are:

> *"Trust is the key to international cooperation" (McGillivray and Smith, 2000).*
>
> *"If negative peace and its three subcategories (fragile, unstable, and cold peace) are defined by the absence of war, then positive peace is defined by the presence of confidence and trust … there seems to be agreement that alliances and international treaties, at least by themselves, provide at best only thin grounds for true and long-lasting mutual understanding … the key is to be found in what more identifies as missing from alliances—mutual confidence and trust" (Oelsner, 2007).*
>
> *"Trust can change over time … when the states' national security interests are at stake and their chain of considerations becomes more selfish and focused on protecting their own interest." (Kallberg, 2012).*

Computer science literatures rely on the political and social sciences literatures for the trust relation definitions. The only difference that, researchers in the computer science (or applied mathematics) try to represent (or formalize) the ambiguousness of *"trust"* and *"relation"* definitions in nature languages into a scientific and practical approach for a computation. A formalism of the relation computation is rooted to probability studies. In earlier stages, Bayes's probability and Dempster-Shaffer theory (Shafer, 1976) were used in belief function for the relation computation. Later, a subjective logic (Jøsang, 2013, 2001) is introduced as a formalism for the relation computation in computer science. An expansion to the subjective logic was offered by (Trček, 2009) for a trust modeling (e.g. trust operators) in computer networks such as trust management in web services, social networks, e-voting, etc. As a conclusion to the trust definitions in the computer science, computer scientists do not define or redefine the trust definitions, it provides a methodology (Trček, 2009) for the relation computation and its statistical inferences.

*Trust Modeling*

Referring to the previous section, there are many trust definitions that were expressed in natural languages. To enable relation computation in computing systems, the trust definitions need to be conveyed in a measurable or quantifiable notations such as statistical representations. The first attempt to represent trust in the statistical notations was by (Dempster, 1967). Dempster showed a probability measurement that is to define an upper and lower probabilities for a multivalued mapping. The probability measurement is a generalization of calculus in Bayesian theory. His statistical scheme is adopted by (Shafer, 1979, 1976) and it provides elegant method to compute trust. Many researchers later (in 1980-1995) addressed both works as the foundation for a concrete relation computation, which they began to call as Dempster-Shafer theory in the early 1980s (Shafer, 2015). Later (Jøsang, 1997) provided an extension to a probabilistic calculus for the Dempster-Shaper theory by introducing an artificial reasoning, named subjective logic. The following subsections will further discuss the Bayesian theory, Dempster-Shafer theory and subjective logic.

*Bayesian Theory*

The Bayesian theory is widely used in probability studies such as statistical inference. The statistical inference is a process to deduce a conclusion from given hypotheses using data that were sampled from a population. Through the statistical inference, a probability of the hypotheses to be either true, false or unknown can be derived using Bayesian probability. The Bayesian probability can be computed using Bayesian inference (or rule). The Bayesian inference allows an update to a probability of a given hypothesis when a new evidence was found. It is contributed to a non-monotonic logic (Brewka, 1992) that allows a tentative conclusion (previously deduced conclusion) to be retracted when the new evidence invalidates the tentative conclusion. This formal logic is used in an artificial intelligent (AI) study for a decision making when past experiences and new experiences are accumulated. In the AI literature, this work is called *belief revision* (Wobcke, 1995). Below is the Bayes's theorem that is used as an inference rule for belief computation as well as belief update.

$$P(H|D) = \frac{P(D|H)\, P(H)}{P(D)}$$

Where P(H|D) is a posterior probability [1] of hypothesis H after datum D is observed. Let $P(H)$ and $P(D)$ be probabilities of two separate events without regard to each other. $P(H)$ is a prior probability of hypothesis H before datum D is observed. P(D|H) is a probability of datum D given that hypothesis H is true.

*Dempster-Shafer Theory*

The theory is also known as the theory of belief function. In the Bayesian theory, each question of interests (or events) requires probabilities that are assigned to each of them. The belief function (Shafer, 1976) measures a *degree of belief* (or *mass*) for one question using subjective probability for a related question. The degree of belief may not bound to mathematical probabilities and its probabilities depending on how closely questions (or events) are related. The belief for one question can be combined using Dempster's rule that when the question relies on independent items of evidence.

The belief function will assign a mass for every hypothesis (or question of interests). A total mass is a measurement of the entire evidence (or a belief of given hypothesis H). The total mass that supports the given hypothesis H will form a lower bound of belief. Plausibility ($Pl$) is a total mass measurement of the entire evidence that contradicts to the given hypothesis H. The plausibility is an upper bound of the given hypothesis H would be true. Let:

$e_H, e_{\neg H} \subseteq E$, s.t. $E$ is a set of all evidence
$e_H$ total evidence of the given hypothesis H
$e_{\neg H}$ total evidence that contradicts the given hypothesis H
$Bel(H) = \sum e_H$ is a total mass of belief H.
$Pl(H) = 1 - \sum e_{\neg H}$ is a plausibility of belief H.

Example 1: An agent A observes a relation between two nations. Given two hypotheses:
$e_{H1} = 15\%, e_{H2} = 5\%\ for\ FRIENDLY$
$e_{H1} = 20\%, e_{H2} = 25\%, e_{H3} = 5\%\ for\ HOSTILE$

Table 1: Relations between two nations

| Hypothesis | Mass | Belief | Plausibility |
|---|---|---|---|
| NULL (Neither FRIENDLY or HOSTILE) | 0% | 0% | 0% |
| FRIENDLY | 15%, 5% | 20% | 50% |
| HOSTILE | 20%, 25%, 5% | 50% | 80% |
| EITHER (FRIENDLY or HOSTILE) | 30% | 100% | 100% |

---

[1] Prior probability: probability that an observation will fall into hypotheses *before data collection*. Posterior probability (or conditional probability): probability of assigning observations to the hypotheses *by given collected data (e.g. new evidences)*.

From Table 1, belief is a total mass (e.g. belief for friendly hypothesis is 20% = 15% + 5%). Plausibility for friendly hypothesis is 50% = 100% − belief not friendly (or hostile 50%) and hostile hypothesis is 80% = 100% − belief not hostile (or friendly 20%). The Null hypothesis is always 0% because it has no solution. Mass for either (hostile or friendly) hypothesis is a gap (or uncertainty) between mass friendly and mass hostile such that 30% = 100% − ((15% + 5%) + (20% + 25% + 5%)). Belief and plausibility are always 100% because the universality of the hypothesis either (friendly or hostile) and this hypothesis is always true in this case.

*Subjective Logic (Algebra)*

The subjective logic (Jøsang, 2015) is used for modeling and analyzing of incomplete information that involves uncertainty in a belief reasoning calculus. Each belief is represented as a collection of opinions in a finite state space[2]. The opinions may contain a degree of uncertainty about its probability (Jøsang, 2013). The degree of uncertainty can be interpreted as ignorance about the truth of a given state (or actual probability). The calculus for the subjective logic consist of an ordered quadruple $\omega_x = (b, d, u, a)$ as an opinion proposition such that:

**belief** $b$, is a belief mass that a proposition is true.
**disbelief** $d$, is a belief mass that a proposition is false.
**uncertainty** $u$, is a belief mass that an unknown neither true nor false of a given proposition.
**base rate** $a$, is a priori[3] probability with the absence of evidence.
Where:
$b, d, u, a \in [0 \dots 1]$, s.t. $b + d + u = 1$
When:
$b = 1$, is a binary logic TRUE.
$d = 1$, is a binary logic FALSE
$b + d = 1$, is a traditional probability (e.g. $b + \neg b = 1$, s.t. $\neg b$ is $d$ using complement)
$b + d < 1$, is a degree of uncertainty.
$b + d = 0$, is a complete uncertainty.
$E_x = b + au$, is a probability projection for $P_x$.

Binomial opinions of triple $(b, d, u)$ can be presented as graphical representation using an equilateral triangle diagram. The binomial and multiple of binomial opinions (or multinomial) can be computed using subjective logic operators such as complement (NOT), addition (ADD), subtraction (SUB), Comultiplication (OR), etc.

*Related Literatures*

(McGillivray and Smith, 2000) discussed an international cooperation that can reflect trust relationships between nations. The authors have argued that a punishment for those defect the international cooperation can be used to make partner nations become more honest and trustworthy. (Mcknight and Chervany, 2001, 1996) revisited a trust and distrust definitions where they collected numerous trust definitions across multiple disciplines since 1970. Based on their findings, they proposed two topologies to represent trust: i) a classification system to determine types of trust; and ii) a trust model that can be derived from definitions of six related trust types (e.g. trusting intention, trusting behavior, trusting beliefs, system trust, disposition trust, and situational and decisional to trust). (Jøsang, 2001) proposed a new opinion metric to represent Dempster-Shafer's belief model with a demonstration of uncertainty and its probability. The author has introduced a subjective logic to represent the opinion metric using standard and non-standard logical operators (e.g. consensus operator that is used for combining evidence between two parties opinion). (Newton, 2001) explored an empirical relationship between a social and political trust. The author's suggestion is to evaluate trust at a society level with consideration of external world factors rather than just an individual level. A performance of government (e.g. corruption, economy, crime, unemployment) will indicate its trustworthiness to a society because it will affect everyone. Therefore, to evaluate the political trust; trust assessments and aggregation analyzes should be done at the society level.

---

[2] In the original literature, it was called a frame of discernment. In this work, we have simplified many notations for ease of reader to grasp the basic ideas of the original literature.

[3] A priori probability is a probability that is deduced from current evidence without concerning other factors that may directly or indirectly affect a given proposition such as perceptions, past experiences, insight, etc. The probability is presumed in the absence of further evidence.

(Kydd, 2005) debated about roles of trust and mistrust in international relations as well as its effects in the Cold War. The author has shortened a trust definition as a belief that another entity is trustworthy and willing to establish cooperation; and mistrust is a belief that another entity is untrustworthy with the intention to exploit the cooperation. A nation with a sufficient trust can increase a positive relation with other nations and it can also benefit to reduce conflicts. To enable the cooperation, each participant entity must exceed a minimal trust threshold that allows the participant entities to begin the cooperation. The author has showed that the minimal trust threshold for the trust and mistrust in the international relations can be evaluated using game theory (e.g. prisoner's dilemma). An example of security dilemma by the authors:

> "Insecure states will pursue power to make themselves more secure; this renders other states less secure, and their efforts to catch up in turn render the first state less secure in a vicious circle. International conflict is a tragic clash between states with fundamentally benign desires to survive."

(Golbeck, 2006) defined a set of properties that can be used to express trust relationships in social networks. (Oelsner, 2007) criticized that a regional peace is not enough to enable regional friendships. The author has categorized the regional peace into two parts: negative peace (e.g. fragile peace, unstable peace, cold or conditional peace) and positive peace (e.g. stable peace and pluralistic security community). Some of trust and distrust properties were stated in their works. The properties will affect the international nation relationships such as recent war, deployment troop in a border, diplomatic visit, antipathy between societies, arms races, join project, people mobility nuclear program and treaty, economic cooperation, etc. (Martinelli and Petrocchi, 2007) defined an integrated framework for modeling of security and trust. (Yan and Holtmanns, 2007) discussed a transformation of social trust into a digital trust that can be used in trust modeling and trust management. (Walter et al., 2008) presented a trust model using recommendation system in a social network. The authors have used multi-agents to find a trust relationship using friend's recommendations. The trust agents were used to keep track of trust values in a communication. (Biskup et al., 2008) proposed trust modalities that are based on correlations between certified properties, experiences and expectations in a control and access decision. (Trček, 2008) summarized collection trust definitions by the previous literature. The most remarkable trust definition that is related to this research work: *"Trust is a subjective probability by which an individual expects that another individual performs..."* (reliability trust) and *"Trust is the extent to which one party is willing to depend on something or somebody in given situation ... even though negative consequences are possible."* (decision trust). (Trček, 2009) proposed a formalism of trust model using a taxonomy of generic model and a qualitative algebra. The qualitative algebra is used for model definitions and to transform it into algebraic forms (formal representation). It includes basic quantitative methodologies for modeling trust relations such as *trusted, untrusted* and *undecided*. The authors have showed that the trust relations can be modeled based on these properties: not reflexive, not symmetric, non-transitive and commutative (for fully connected society). The proposed trust model was tested in web services for an automated trust management applications (Jelenc and Trček, 2014; Trček, 2011).

Kallberg (Kallberg, 2012) identified trust as an element that is important to ensure that members of Common Criteria are able to recognize and consume Common Criteria's certified products and services. He has argued that "*long-term survival of CC requires abandoning the global approach and instead use established groupings of trust*". His major suggestion is to use a customized group of Common Criteria that based on mutual interest such as defense alliance, economic cooperation agreements, historical events, and political alliances; because it conveys transitive trust between its partners. The author has viewed the Common Criteria concerns from the perspective of relationships and trust boundaries between nations, which he considered as major issues. He has proposed *"group of trust"* as a trivial solution for these problems (Koster et al., 2012) proposed a trust alignment that can be used for finding a translation of the other agent's trust evaluations using shared evidence. The authors' assumption that a shared information can be observed by participant agents. (J.Wheeler, 2012) studied challenges and clarifications in building trust relationships for international relations when conflict situations happened. The author has discussed diversity trust relationships from the origin of trust definitions, identification of mistrust issues in the international relations, and the challenges that are to sustain and improve the trust relationships. An example of internal issues regarding civil nuclear technology between India and Pakistan (1999-2002) was also discussed that is either for peaceful or militarization (defensive to offensive) purpose – nuclear dilemmas for leaders' interpretation and response between India and Pakistan. He also mentioned that an ideology conviction for a nation can imply bad faith thinking for trust-building to another nation (e.g. Cold War: US (capitalism, liberal democracy) and the Soviet Union (socialism, communist Marxism–Leninism, totalitarianism). The ideology conviction was also to be claimed to be the cause of bad relations between India-Pakistan, Israel-Palestine (Eldad, 2013) and Iran-US. (Rohner et al., 2012) proposed a theory that is to study trust and conflict with correspondence to an international trade. The authors' argument that, business relationships (e.g. international trade) will be disrupted if two nations struggle in soft or hard conflicts. The conflict will render trust and it will cause scant trade between the struggled nations. The authors' suggestion that, an inverse relation such that if we are increase trust in trader communities by means of foster the trade activities, then it will decrease the conflict between the struggled nations. The suggestion is consistent with a well-known theory (e.g. rational-agent theory) and its prediction of a negative correlation between war and trust.

(Yu et al., 2015) analyzed bilateral trade patterns in 16 European countries between years 1996 to 2009. The authors found that when there is an imbalance of importing and exporting rule and regulations between importer country and exporter country, the effect of trust on trade will be changed consequently. For example, a tight importing rule and regulations will reduce the trust of the exporter country toward to the importing country. (Bansal and Mariam, 2015) studied the effect of privacy violation and its correlation to a trust violation and repair in a security breach. The authors have explored the security breach problem such as leaking of users' privacy data and how an information security officer that responsible to the security breach will response to the user. (Fang et al., 2015) addressed multi-facets of trust issues in social sciences (Svetlana and Tzukanova, 2016). The authors' suggestion is to use a distrust framework to address the multi-facets of trust issues. The distrust framework will cover an interpersonal (e.g. benevolence, competence, integrity and predictability) and impersonal (e.g. user connectivity in trust networks) issues.

**Methodology**

In this section, we present a brief summary of our research works. We have begun the research works by constructing research hypotheses. We derived our research hypotheses by literature reviews that are related to trust and international relations. During the literature reviews phase, we have identified case studies for the relation algebra (in Case Study section). Based on the case studies, we collected data that are needed to compute the relation algebra for the international relations. The outcomes of the research works that are a method to compute trust between nations (or trust value) wherein a trust perception is derived from the trust value.

**Relation Algebra: Definition and Notation**

In this section, we present the relation algebra and its examples.

**DEFINITION 1.** A nation state is a sovereign nation and recognized by the United Nation (UN). Referring to the UN's Charter ("Charter of the United Nations," 1945):

Chapter I, Articles 1: "*To maintain international peace and security…*" and "*to develop friendly relations among nations based on respect for the principle of equal rights…*". Articles 2: "*…principle of the sovereign equality of all its Members.*".

Chapter II, Articles 4: "*Membership in the United Nations is open to all other peace-loving states which accept the obligations contained in the present Charter … and, in the judgment of the Organization, are able to carry out these obligations*".

Referred to the UN's Charter, we defined a *nation* term as the nation state or any UN member states.

**DEFINITION 2.** A relation is international relations between Nation A and Nation B. The relation can be either friendly (ally, positive), neutral, or hostile (enemy, negative). The relation $\mathcal{R}_{A,B}$ denotes a *trust perception* of Nation A toward Nation B. Let assume that:

$RELATION$ and $NATION$ are sets
$f \stackrel{\text{def}}{=} friendly,\ n \stackrel{\text{def}}{=} neutral,\ h \stackrel{\text{def}}{=} hostile$
$f, n, h \in RELATION$
$A, B \in NATION$
$\mathcal{R}_{A,B} = (f \cap n \cap h) = \emptyset$
$\mathcal{R}_{A,B} = (f \cap n) \cup (f \cap h) \cup (n \cap h) = \emptyset;$

**Remark 2.1.** The relation for $\mathcal{R}_{A,A}$ is reflexive with always friendly. We assumed that a nation always trust itself.

**Remark 2.2.** The relation for $\mathcal{R}_{A,B} \neq \mathcal{R}_{B,A}$ is not always symmetric. In this thesis, we assumed that $\mathcal{R}_{A,B} = \mathcal{R}_{B,A}$ is symmetric, which is to simplify the modeling work.

**Remark 2.3.** Relation for $\mathcal{R}_{A,B}$ and $\mathcal{R}_{B,C}$ does not always imply that $\mathcal{R}_{A,C}$ is transitive for relations between Nation A, Nation B and Nation C.

**Remark 2.4.** The relation for $\mathcal{R}_{A,B}$ and $\mathcal{R}_{B,A}$ are commutative for binary operation $(\mathcal{R}_{A,B}, \mathcal{R}_{B,A}) = (\mathcal{R}_{B,A}, \mathcal{R}_{A,B})$ for addition and multiplication operations.

**DEFINITION 3.** A relation for Nation A and Nation B is undefined for $\boldsymbol{\mathcal{R}_{A,B} = (f \cup n \cup h) = \emptyset}$.

**Remark 3.1.** The relation for $\mathcal{R}_{A,B}$ is undefined when the relation between Nation A and Nation B is neither friendly, neutral, nor hostile. The state of the relation is unknown.

**Remark 3.2.** If a definition of a *nation* is reduced to **DEF 1** then international relations always exists because of diplomatic relations and recognitions. Later we will show that the undefined relation (total uncertainty) does not exist in the real world except in a mathematical form.

**DEFINITION 4.** A weightage is used for a linear normalization of trust perception between Nation A toward Nation B. The weightage will help to identify the significance the trust perception. It is almost similar to a base rate in a subjective logic.

**THEOREM 1.** Mass Weightage

Let assume that:

$x \in \mathbb{Z}, \quad x \geq 1;$

$\mathcal{C} = |RELATION|$ or $cardinality(RELATION)$

Mapped matrix: $\boldsymbol{RELATION} \times (1 \leq X \leq \mathcal{C})^T$

$[f \quad n \quad h] \times \begin{bmatrix} 1 \\ 2 \\ 3 \end{bmatrix} \; s.t. (f \mapsto 1), (n \mapsto 2) \; and \; (h \mapsto 3).$

$$\mathcal{W}_{Mass} = \sum_{x \leq \mathcal{C}}^{x=1} \mathcal{W}_x = 1, \quad \mathcal{W}_x \in \mathbb{R}, \quad 0 \leq \mathcal{W}_x \leq 1$$

**Remark 4.1.** One may choose to use a priori probability to evaluate (assign value) for every $\mathcal{W}_x$. Given that a cardinality is equal to three, then each $\mathcal{W}_x$ is equal to $\frac{1}{3}$. One may also use a different value of $\mathcal{W}_x$ that is based on the number of properties $\mathcal{P}$ as mentioned in Theorem 3. For large numbers of the properties $\mathcal{P}$ for a given $\mathcal{W}_x$, the $\mathcal{W}_x$ should be increased to represent large samples of the properties $\mathcal{P}$. However, the value of $\mathcal{W}_x$ is subjective to an observer. For example, if the observer wants to see differences between hostile and friendly relations for $\mathcal{R}_{A,B}$, then a neutral $\mathcal{W}_{n \mapsto 2}$ should be decreased; while a friendly $\mathcal{W}_{f \mapsto 1}$ and hostile $\mathcal{W}_{h \mapsto 3}$ should be increased.

**DEFINITION 5.** A scalar determines an interval scale for international relations, which is either friendly, neutral, or hostile.

**THEOREM 2.** Mass Scalar

Let assume that:

$\mathcal{S}_1 = h's \; scalar \; sign = -1$
$\mathcal{S}_2 = n's \; scalar \; sign = +1$
$\mathcal{S}_3 = f's \; scalar \; sign = +1$

One may choose scalar signs: either "+" or "−")[4].

$$\mathcal{S}_{Mass} = \sum_{x \leq \mathcal{C}}^{x=1} |\mathcal{S}_x . \mathcal{W}_x| = 1$$

**Lemma 2.1.** Lower bound, middle bound and upper bound in Mass Scalar (interval scale).

$\mathcal{S}_{Lower} = \mathcal{S}_1 . \mathcal{W}_1$

$$\mathcal{S}_{upper} = \sum_{x \leq \mathcal{C}}^{x=2} \mathcal{S}_x . \mathcal{W}_x$$

$\mathcal{S}_{Lower} + \mathcal{W}_1 \leq \mathcal{S}_{middle} \leq (\mathcal{S}_{upper} - \mathcal{S}_\mathcal{C} . \mathcal{W}_\mathcal{C})$

**DEFINITION 6.** Perception is a collection of relation properties or elements that are used to determine a relation alignment for $\mathcal{R}_{A,B}$. The relation properties $\mathcal{P}$ are nominal data that assigned with some value based on qualitative statistics or by an observer intuition.

---

[4] We chose to use a negative sign for a hostile and positive sign for a neutral and friendly relations. By common sense, the negative sign may suitable to be used for the hostile relation.

**Remark 6.1** Relations $\mathcal{R}_{A,B}$ will have a collection of relation properties $\mathcal{P}$ for each relation perception (e.g. $f_\mathcal{P}, n_p, h_p$). Each trust property $\mathcal{P}_x$ can be mapped into nominal data with values such as war ally $\mathcal{P}_1 = 0.2$, war enemy $\mathcal{P}_2 = 0.2$, politic $\mathcal{P}_1 = 0.075$, trade $\mathcal{P}_1 = 0.1$, spy and counter intelligent $\mathcal{P}_1 = 0.15$ etc.

**THEOREM 3.** Mass Properties

Assume that:

$\mathcal{P}_x \in \mathbb{R}, \quad s.t.\ 0 \leq \mathcal{P}_x \leq 1;$

Let cardinalities: $i, j, k$

$i = |friendly\ properties|$
$j = |neutral\ properties|$
$k = |hostile\ properties|$

$$h_{\mathcal{P}_{Mass}} = \sum_{x=k}^{x=1} \mathcal{P}_x \leq 1$$

$$n_{\mathcal{P}_{Mass}} = \sum_{x=j}^{x=1} \mathcal{P}_x \leq 1$$

$$f_{\mathcal{P}_{Mass}} = \sum_{x=i}^{x=1} \mathcal{P}_x \leq 1$$

**DEFINITION 7.** Relation $\mathcal{R}_{A,B}$ is a product of Mass Perception $\mathcal{T}_{Mass}$. The Mass Perception is a point that resides in a relative distance between a lower bound and upper bound of Mass Scalar. To determine international relations for $\mathcal{R}_{A,B}$, which either friendly, neutral or hostile:

- If the point falls into less than middle bound, it is a hostile relation;
- If the point falls into greater than middle bound, it is a friendly relation; and
- If the point falls into the middle bound, it is a neutral relation.

**Remark 7.1.** Theorems 1 through 4 rely on three major conditions of international relations, which are hostile, neutral and friendly. One may define more than triple conditions to implement granularity and fuzziness in the international relations. For example, a septuple with additional three relations such as Near-Hostile, Near-Neutral and Near-Friendly as shown in Fig. 1 can be used.

**Remark 7.2.** One should not modify the triple conditions to implement additional relations because it will increase difficulties in properties $\mathcal{P}$ classification and nominal data (value assignment). For an example, to identify and assign nominal values for Hostile's properties is much easier that compare to doing a similar work for Near-Hostile's properties. These will increase statistical efforts for data collection, data interpretation, data analysis, data computation etc. We usually apply qualitative methods for international relations (Detlef F. Sprinz, 2004; Pollack, 2001; Walt, 1998) and most data that are in raw forms (e.g. plaintexts in news, books, research publications, government official portal, documentary, online media, hacked/leaked classified information in public domains[5] etc.), and some data (e.g. classified data) are not always available on the Internet because it was protected by government[6]. As a suggestion for more than triple scalar of relations, that is to directly map the Mass Perception's value in the Theorem 4 into the septuple scalar as shown in Fig. 1. One must define a lower bound and an upper bound for each new relation element. The new relation element is a subset of the existing triple (e.g. Near-Hostile $\subset$ Hostile).

**THEOREM 4.** Mass Trust Perception

Mapped matrix **Mass Properties** $\times$ **Scalar**

---

[5] *In this research work, we do not obtain or consume any material that may lead to actions of a cyber-criminal, terrorism, spying activities and any other illegal activities.*

[6] *An attempt to obtain and consume the classified information may lead to a cyber-crime (or spying) activity – it may happen unintentionally.*

$$\mathcal{T}_{Mass} = \sum( [f_{\mathcal{P}_{Mass}} \quad n_{\mathcal{P}_{Mass}} \quad h_{\mathcal{P}_{Mass}}] \times \begin{bmatrix} \mathcal{S}_1.\mathcal{W}_1 \\ \mathcal{S}_2.\mathcal{W}_2 \\ \mathcal{S}_3.\mathcal{W}_3 \end{bmatrix} )$$

$$\mathcal{R}_{A,B}(\mathcal{T}_{Mass}) = \begin{cases} hostile, & \mathcal{T}_{Mass} < \mathcal{S}_{Lower} - \mathcal{W}_1 \\ neutral, & \mathcal{T}_{Mass} = \mathcal{S}_{middle} \\ friendly, & \mathcal{T}_{Mass} > \mathcal{S}_{upper} - \mathcal{S}_\mathcal{C}.\mathcal{W}_\mathcal{C} \end{cases}$$

**Lemma 4.1.** Strength of Mass Perception

$$\mathcal{T}_{Mass\ Strength} = \sum( [f_{\mathcal{P}_{Mass}} \quad n_{\mathcal{P}_{Mass}} \quad h_{\mathcal{P}_{Mass}}] \times \begin{bmatrix} \mathcal{W}_1 \\ \mathcal{W}_2 \\ \mathcal{W}_3 \end{bmatrix} )$$

**Remark 7.3.** The $\mathcal{T}_{Mass\ Strength}$ is a total of all masses without applying negative value (not apply negative scalar sign) to hostile properties. The purpose of the $\mathcal{T}_{Mass\ Strength}$ is to evaluate relations (or enabled properties) between hostile properties, and neutral and friendly properties. It used as a probability function (PF) wherein $\mathcal{T}_{Mass\ Strength}$ is conserved as the total probability for all enabled masses in given relations. The following paragraph discuss some examples of interpretations between $\mathcal{T}_{Mass}$ and $\mathcal{T}_{Mass\ Strength}$.

When $\mathcal{T}_{Mass\ Strength}$ is near to 1, $\mathcal{T}_{Mass}$ may represent many contradiction of opinions between relations hostile and friendly. This may happen if it involves a long duration of sampling (or observation) of international relations between two nations. If the $\mathcal{T}_{Mass}$ is a product of a 15-year observation period for the international relations between two nations: it may consist of a year of war, a year of military ally, a year of politics disagreement, a year of economy sanction etc. If the $\mathcal{T}_{Mass}$ is a product of less than a 5-years observation period, the contradiction of opinions may occur when a nation leader or ruling party (Bansal and Mariam, 2015; Heitmeyer, 2009; Menon et al., 2014; Pollack, 2001; Walt, 1998) was changed due to an election, revolution, installation of puppet leader (e.g. Karzai-Afghanistan (Rod Nordland, 2014)) as a post-war outcome etc.

When $\mathcal{T}_{Mass\ Strength}$ is near to 0.5 (or middle), $\mathcal{T}_{Mass}$ may represent a fair opinion that either relation hostile or friendly. If the $\mathcal{T}_{Mass}$ is a product of observation for many years (e.g. 15 years), it may represent consistent international relations in that duration. When $\mathcal{T}_{Mass\ Strength}$ is near to $n_{\mathcal{P}_{Mass}}$, $\mathcal{T}_{Mass}$ represents a bias to a relation neutral. If the $\mathcal{T}_{Mass}$ is a product of observation for many years, it may represent a firm of relation neutral at that moment. When $\mathcal{T}_{Mass\ Strength}$ and $n_{\mathcal{P}_{Mass}}$ are identical in a positive value, it indicates that there are no hostile properties in the calculation (or observation).

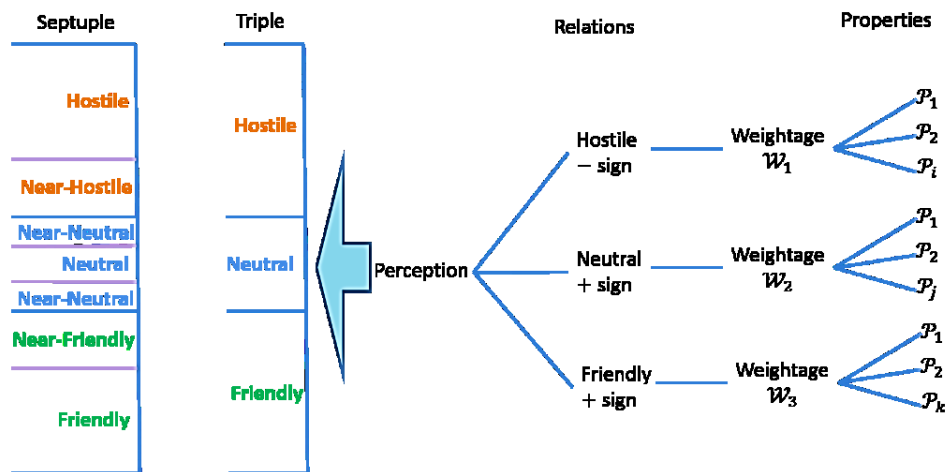

Fig. 1. A summary of algebra relation for international relations

Table 2: An Example of Relation Computation

| Relation | Hostile | Neutral | Friendly |
|---|---|---|---|
| Properties | $\mathcal{P}_1 = 0.5$ | $\mathcal{P}_1 = 0.5$ | $\mathcal{P}_1 = 0.05$ |

| | $\mathcal{P}_x$ | $\mathcal{P}_2 = 0.3$ | $\mathcal{P}_2 = 0$ | $\mathcal{P}_2 = 0$ |
|---|---|---|---|---|
| | | $\mathcal{P}_3 = 0$ | $\mathcal{P}_3 = 0$ | $\mathcal{P}_3 = 0$ |
| | | $\mathcal{P}_4 = 0.15$ | $\mathcal{P}_4 = 0.1$ | $\mathcal{P}_4 = 0$ |
| | | $\mathcal{P}_5 = 0.05$ | | $\mathcal{P}_5 = 0$ |
| | | $\mathcal{P}_6 = 0$ | | $\mathcal{P}_6 = 0.1$ |
| | Weightage $\mathcal{W}_{Mass}$ | 0.45 | 0.10 | 0.45 |
| | Scalar Sign $\mathcal{S}_x$ | $\mathcal{S}_1$ is $-$ | $\mathcal{S}_2$ is $+$ | $\mathcal{S}_3$ is $+$ |

$h_{\mathcal{P}_{Mass}} = 0.9 = 0.5 + 0.2 + 0.15 + 0.05$

$n_{\mathcal{P}_{Mass}} = 0.6 = 0.5 + 0.1$

$f_{\mathcal{P}_{Mass}} = 0.15 = 0.05 + 0.1$

$\mathcal{S}_{Mass} = 1 = 0.45 + 0.10 + 0.45$

$\mathcal{S}_{Lower} = -0.45$

$\mathcal{S}_{upper} = 0.55 = 0.10 + 0.45$

$-0.45 + 0.45 \leq \mathcal{S}_{middle} \leq (0.55 - 0.45)$

$0 \leq \mathcal{S}_{middle} \leq 0.10$

$\mathcal{T}_{Mass} = \sum( [0.9 \quad 0.6 \quad 0.15] \times \begin{bmatrix} -.0.45 \\ +.0.10 \\ +.0.45 \end{bmatrix} )$

$\mathcal{T}_{Mass} = -0.2775 = \sum( [-0.405 \quad +0.06 \quad +0.0675])$

$\mathcal{R}_{A,B}(-0.2775) = \begin{cases} \textbf{hostile,} & \mathcal{T}_{Mass} \text{ is } < S\_middle \\ \textbf{neutral,} & \mathcal{T}_{Mass} \text{ is } S\_middle \\ \textbf{friendly,} & \mathcal{T}_{Mass} \text{ is } > S\_middle \end{cases}$

$\mathcal{T}_{Mass\ Stength} = 0.5325 = \sum( [0.9 \quad 0.6 \quad 0.15] \times \begin{bmatrix} 0.45 \\ 0.10 \\ 0.45 \end{bmatrix} )$

Referring to Theorem 4 and Lemma 4.1, the given example has shown that a relation between Nation A and Nation B is hostile. The strength of the relation is near to 0.5 such that it represents a consistent hostile relation during the observation.

**Case Study: International Relations**

In this section, we explore international nation relations between the USA-GBR, USA-IRN and USA-IND. We deliberated about the properties that were associated with international events. We also discuss some of the problems encountered during data collections and testing of the relation algebra with the given case studies.

*Properties*

We have clustered events that may affect international nation relations as showed in Tables 3, 4 and 5. Clustering or grouping the related events for certain properties will reduce complexities for determining properties' values and it will help to reduce the searching time of the whole data in public domains (e.g. Internet, news, etc.). If at least a single event is found to be related to the given properties, then the given properties will be included in a relation computation. It may not be strong enough as a solid evidence for the given properties, but it will help to enable the relation computation. The Dempster-Shafer's theory of evidence may also be applied in event verifications. However, it requires too much effort.

Another issue with the properties, which one or more events that were assigned to properties may enable and disable (or toggle multiple times) during the observation period. For an example, in neutral's properties, a diplomatic mission (or embassy) may open during a negotiation is still acceptable but later it was closed due to relation crisis – how do we evaluate this toggled event? Should the observer choose to enable or disable the properties, or enable the properties by dividing it into half of the properties' value?

Table 3. Friendly (Positive)

| $\mathcal{P}_x$ | Descriptions |
|---|---|
| $\mathcal{P}_1$ 0.5 | War ally and mutual defense pact during war. |
| $\mathcal{P}_2$ 0.2 | Share/trade nuclear technologies and materials (e.g. uranium) or mass destruction weapon for warfare. |

| $\mathcal{P}_x$ | Description |
|---|---|
| | Arm collaboration in R&D for warfare. |
| | Financial aid for warfare. |
| $\mathcal{P}_3$ 0.1 | Head of the state political sentiment and relationships. |
| $\mathcal{P}_4$ 0.1 | Loan or share strategic technologies and equipment. |
| | Civil nuclear trade and agreement. |
| | Defense pact that enable during peace. |
| $\mathcal{P}_5$ 0.075 | Share military intelligent. |
| | Large scale of joint military drills. |
| $\mathcal{P}_6$ 0.025 | Global War on Terrorism (GWOT) |
| 1.0 | TOTAL |

Table 4. Neutral

| $\mathcal{P}_x$ | Descriptions |
|---|---|
| $\mathcal{P}_1$ 0.25 | Member of UN or nation state recognized by UN. |
| $\mathcal{P}_2$ 0.35 | Economic cooperation. E.g. Bilateral trade, multilateral open market, free trade. |
| $\mathcal{P}_3$ 0.40 | Diplomatic mission (embassy or representative). |
| | Disaster aid and peacekeeping. |
| 1.0 | TOTAL |

Table 5. Hostile (Negative)

| $\mathcal{P}_x$ | Descriptions |
|---|---|
| $\mathcal{P}_1$ 0.5 | War Enemy |
| $\mathcal{P}_2$ 0.2 | Strong disapproval of share/trade/usage nuclear technologies and materials, or mass destruction weapon. E.g. nuclear testing, intercontinental ballistic missile (ICBM) development and testing, and arms races. |
| $\mathcal{P}_3$ 0.075 | Economy blockage or sanction. |
| | Embargo or boycott. (e.g. large scale product boycott, ban visa) |
| $\mathcal{P}_4$ 0.125 | Closed border military aggressive or hostility. Including land, air, maritime trespassing and terrorism. |
| | *peaceful dispute through international law is not included. |
| $\mathcal{P}_5$ 0.05 | Political sentiments and threat by the head of state. |
| $\mathcal{P}_6$ 0.05 | Kill or arrest another nation diplomats. |
| | Espionage. (e.g. spying and hacking) |
| 1.0 | TOTAL |

*Weightage*

We chose to implement 40%:20%:40% as weightages for hostile, neutral and friendly relations. The weightage percentages were decided based on the number of properties for the given relations.

**Case 1: The USA and GBR (1999-2014)**

The United States of America and Great Britain enjoy a long lasting of good international relations (Great Britain. Parliament. House of Commons. Foreign Affairs Committee, 2010). The British-America (or Anglo-American) relation remains intact as a close military ally since the World War II. Both nations also share many strategic resources (Great Britain. Parliament. House of Commons. Foreign Affairs Committee, 2010) and information (e.g. UKUSA Agreement (NSA, 2013) ).

Table 6. Properties of the USA and GBR

| $\mathcal{R}_{USA,GBR}$ | Hostile | Neutral | Friendly |
|---|---|---|---|
| Properties $\mathcal{P}_x$ | $\mathcal{P}_1 = 0$ | $\mathcal{P}_1 = 0.25$ | $\mathcal{P}_1 = 0.5$ |
| | $\mathcal{P}_2 = 0$ | $\mathcal{P}_2 = 0.35$ | $\mathcal{P}_2 = 0.2$ |
| | $\mathcal{P}_3 = 0$ | $\mathcal{P}_3 = 0.40$ | $\mathcal{P}_3 = 0.05$ |
| | $\mathcal{P}_4 = 0$ | | $\mathcal{P}_4 = 0.125$ |
| | $\mathcal{P}_5 = 0$ | | $\mathcal{P}_5 = 0.075$ |
| | $\mathcal{P}_6 = 0$ | | $\mathcal{P}_6 = 0.05$ |
| Weightage $\mathcal{W}_{Mass}$ | 0.40 | 0.20 | 0.40 |
| Scalar Sign $\mathcal{S}_x$ | $\mathcal{S}_1$ is − | $\mathcal{S}_2$ is + | $\mathcal{S}_3$ is + |

$h_{\mathcal{P}_{Mass}} = 0.0$
$n_{\mathcal{P}_{Mass}} = 1.0 = 0.25 + 0.35 + .40$
$f_{\mathcal{P}_{Mass}} = 1.0 = 0.5 + 0.2 + 0.05 + 0.125 + 0.075 + 0.05$
$\mathcal{S}_{Mass} = 1$
$\mathcal{S}_{Lower} = -0.4$
$\mathcal{S}_{upper} = 0.6$
$0 \leq \mathcal{S}_{middle} \leq 0.2$
$\quad \mathcal{T}_{Mass} = 0.60 = -0.4 \times 0.0 + 0.2 \times 1.0 + 0.4 \times 1.0$

$\mathcal{R}_{USA,GBR}(0.60) = \begin{cases} hostile, & \mathcal{T}_{Mass} \text{ is} < \mathcal{S}_{middle} \\ neutral, & \mathcal{T}_{Mass} \text{ is } \mathcal{S}_{middle} \\ friendly, & \boldsymbol{\mathcal{T}_{Mass} \text{ is} > \mathcal{S}_{middle}} \end{cases}$

$\mathcal{T}_{Mass\ Strength} = 0.60 = 0.4 \times 0.0 + 0.2 \times 1.0 + 0.4 \times 1.0$

$\mathcal{R}_{USA,GBR}$ showed that the $\mathcal{T}_{Mass}$ and $\mathcal{T}_{Mass\ Strength}$ are identical. When both variables are identical in a positive value, it indicates that there is no hostile property in the observation. The strength of the relations is greater than 0.5, which represent an excellence friendly relations in 1999 until 2014.

**Case 2: The USA and IRN (1999-2014)**

The United States of America and Islamic Republic of Iran did not have a formal diplomatic relation by an ambassador or diplomat. Both nations established indirect diplomatic using other nation embassies, for examples Iran use Pakistan embassy in Washington D.C. ("Interests Section of the Islamic Republic of Iran," 2015) and the USA use Switzerland embassy in Tehran ("Embassy of Switzerland -Foreign Interests Section," 2015). There were many international conflicts happened for both nations during the observation period (Kashani, 2015). The USA saw Iran as a threat to the world peace when Iranian Government began to utilize nuclear energy. Later, Iran was accused of doing a mass destruction weapon development instead of a civil nuclear development (Katzman, 2009). The international relations for both nations were severely strained by the nuclear issue and it elevates to international sanction by the USA (Sherman, 2013; US Department of the Treasury, 2014; Wikipedia, 2015a). After the September 11 attacks on the USA and invasion of Iraq by the USA in 2003, Iran was accused of supporting and exporting terrorism in the world, which included terrorism in Afghanistan and other nations (Mir H. Sadat and Hughes, 2010; "Public Statement on U.S. Policy Toward the Iran Nuclear Negotiations," 2015, "The 'Grand Bargain' Fax: A Missed Oppurtunity," 2002; Sherman, 2013).

Table 7. Properties of the USA and IRN

| $\mathcal{R}_{USA,IRN}$ | Hostile | Neutral | Friendly |
|---|---|---|---|
| Properties $\mathcal{P}_x$ | $\mathcal{P}_1 = 0$ | $\mathcal{P}_1 = 0.25$ | $\mathcal{P}_1 = 0$ |
| | $\mathcal{P}_2 = 0.2$ | $\mathcal{P}_2 = 0.35/2$ | $\mathcal{P}_2 = 0$ |
| | $\mathcal{P}_3 = 0.075$ | $\mathcal{P}_3 = 0$ | $\mathcal{P}_3 = 0$ |
| | $\mathcal{P}_4 = 0.125$ | | $\mathcal{P}_4 = 0$ |
| | $\mathcal{P}_5 = 0.05$ | | $\mathcal{P}_5 = 0$ |
| | $\mathcal{P}_6 = 0.05$ | | $\mathcal{P}_6 = 0$ |
| Weightage $\mathcal{W}_{Mass}$ | 0.40 | 0.20 | 0.40 |
| Scalar Sign $\mathcal{S}_x$ | $\mathcal{S}_1$ is − | $\mathcal{S}_2$ is + | $\mathcal{S}_3$ is + |

$h_{\mathcal{P}_{Mass}} = 0.50 = 0.2 + 0.075 + 0.125 + 0.05 + 0.05$
$n_{\mathcal{P}_{Mass}} = 0.425 = 0.25 + (\frac{0.35}{2})$
$f_{\mathcal{P}_{Mass}} = 0.0$
$\mathcal{S}_{Mass} = 1$
$\mathcal{S}_{Lower} = -0.4$
$\mathcal{S}_{upper} = 0.6$
$0 \leq \mathcal{S}_{middle} \leq 0.2$
$\quad \mathcal{T}_{Mass} = -0.115 = -0.4 \times 0.5 + 0.2 \times 0.425 + 0.4 \times 0.0$

$$\mathcal{R}_{USA,IRN}(-0.115) = \begin{cases} \text{hostile}, & \mathcal{T}_{Mass} \text{ is} < \mathcal{S}_{middle} \\ \text{neutral}, & \mathcal{T}_{Mass} \text{ is } \mathcal{S}_{middle} \\ \text{friendly}, & \mathcal{T}_{Mass} \text{ is} > \mathcal{S}_{middle} \end{cases}$$

$\mathcal{T}_{Mass\ Strength} = 0.285 = 0.4 \times 0.5 + 0.2 \times 0.425 + 0.4 \times 0.0$

$\mathcal{R}_{USA,IRN}$ showed the $\mathcal{T}_{Mass}$ fall into negative value and it is less than $\mathcal{T}_{Mass\ Strength}$. It indicates that there were many hostile's properties in the observation. The strength of the relations is not within $\mathcal{S}_{middle}$ boundaries, which represent a strong of hostility relations in 1999 until 2014. Observe that, there is toggled property in neutral $\mathcal{P}_2 = 0.35/2$, which cause the value is divided by 2. We assumed it as toggled event because the international trade sanction imposed by the USA, but simultaneously the USA is still establishing trade with Iran, which reported by the U.S. Census Bureau (Commerce, 2014).

**Case 3: The USA and IND (1999-2014)**

The United States of America and India relations or Indo-American relations had been improving during the observation period. The USA President Bill Clinton imposed economic sanctions on India because Indian Prime Minister Atal Bihari Vajpayee had authorized nuclear weapon testing at Pokhran (Business Standard, 1998; Dan Balz and William Drozdiak, 1998). *""They clearly create a dangerous new instability in their region and, as a result, in accordance with U.S. law, I have decided to impose economic sanctions against India," Clinton said. Sanctions are mandatory under U.S. law when an undeclared nuclear state explodes a nuclear device. …"In hopes of averting an arms race in southern Asia, specifically in next-door Pakistan, Clinton urged India's neighbors "not to follow the dangerous path India has taken.""*(CNN, 1998).

The relations of both nations improved after the India-United States Civil Nuclear Agreement (or 123 Agreement) signed on 10 October 2008, which allow India to use civil nuclear energy and it also enable civil nuclear trade for both nations (Council on Foreign Relations, 2007; Esther Pan, 2010). India supported President Bush and Obama in a war against terrorism and both nations had waged a war against the Taliban Government in Afghanistan (BBC News, 2001; Wikipedia, 2015b).

Table 8. Properties of the USA and IND

| $\mathcal{R}_{USA,IND}$ | Hostile | Neutral | Friendly |
|---|---|---|---|
| Properties $\mathcal{P}_x$ | $\mathcal{P}_1 = 0$ | $\mathcal{P}_1 = 0.25$ | $\mathcal{P}_1 = 0$ |
| | $\mathcal{P}_2 = 0.2$ | $\mathcal{P}_2 = 0.35/2$ | $\mathcal{P}_2 = 0$ |
| | $\mathcal{P}_3 = 0.075$ | $\mathcal{P}_3 = 0.40$ | $\mathcal{P}_3 = 0$ |
| | $\mathcal{P}_4 = 0$ | | $\mathcal{P}_4 = 0.125$ |
| | $\mathcal{P}_5 = 0$ | | $\mathcal{P}_5 = 0.075$ |
| | $\mathcal{P}_6 = 0.05$ | | $\mathcal{P}_6 = 0.05$ |
| Weightage $\mathcal{W}_{Mass}$ | 0.40 | 0.20 | 0.40 |
| Scalar Sign $\mathcal{S}_x$ | $\mathcal{S}_1$ is $-$ | $\mathcal{S}_2$ is $+$ | $\mathcal{S}_3$ is $+$ |

$h_{\mathcal{P}_{Mass}} = 0.325 = 0.2 + 0.075 + 0.05$
$n_{\mathcal{P}_{Mass}} = 0.825 = 0.25 + 0.35/2 + .40$
$f_{\mathcal{P}_{Mass}} = 0.25 = 0.125 + 0.075 + 0.05$
$\mathcal{S}_{Mass} = 1$
$\mathcal{S}_{Lower} = -0.4$
$\mathcal{S}_{upper} = 0.6$
$0 \leq \mathcal{S}_{middle} \leq 0.2$
$\mathcal{T}_{Mass} = 0.135 = -0.4 \times 0.325 + 0.2 \times 0.825 + 0.4 \times 0.25$

$$\mathcal{R}_{USA,IND}(0.135) = \begin{cases} \text{hostile}, & \mathcal{T}_{Mass} \text{ is} < \mathcal{S}_{middle} \\ \boldsymbol{neutral}, & \boldsymbol{\mathcal{T}_{Mass} \text{ is } \mathcal{S}_{middle}} \\ \text{friendly}, & \mathcal{T}_{Mass} \text{ is} > \mathcal{S}_{middle} \end{cases}$$

$\mathcal{T}_{Mass\ Strength} = 0.395 = 0.4 \times 0.325 + 0.2 \times 0.825 + 0.4 \times 0.25$

$\mathcal{R}_{USA,IND}$ showed the $\mathcal{T}_{Mass}$ and $\mathcal{T}_{Mass\ Strength}$ difference by 0.26. When both variables are identical in a positive value and a gap between both values (0.26) is greater than $\mathcal{S}_{middle}$ (>0.2), it indicates that there are exist hostile and friendly properties in the observation. The strength of the relations is less than 0.5 and the gap is greater than $\mathcal{S}_{middle}$, which represents a fragile neutral relations in 1999 until 2014.

**Results and Discussion**

We have presented case studies for the relation algebra in the Relation Algebra section. Both case studies discussed the international nation relations between USA-GBR (friendly relation), USA-IRN (hostile relation) and USA-IND (neutral relations). The weightage for trust perceptions in the relations are proportions to 2:1:2. The total properties that were considered in the observation are fifteen. Properties identification and properties value assignment are the most difficult part of the relation evaluation process. Clustering or grouping the related events for certain properties will reduce complexities for determining the properties' value. There must be a justification for each selected properties and its values. The justification efforts required an assistance of political and social science experts in the domain. A consul (or officer) in a foreign ministry may not give an honest answer if asked directly for their opinion regarding properties' value and weightages, due to possibly classified information policy concerning its foreign relations. But, if the relation algebra method is used by the foreign ministry or defense ministry, the relation algebra computation will be more accurate because the method can use classified information, which known by the ministry. We can conclude that, the properties and weightages are subjective to the observers. In this work, the properties and weightages chosen by the authors are based on public information available in the literatures (refer to the Literature Review section) and the Internet[7].

The Dempster-Shafer's theory allows multiple events and its observation results (or opinions based on evidences) to be combined to derive a degree of belief or belief function (Shafer, 1976). To test a Dempster's rule, one may use evidence operators based on subjective logic (Jøsang, 2013). It allows multiple agents (observers) to observe for one event at the same time or at difference periods, then all observation results can be combined to draw a tentative conclusion. Additional events and its evidences, and new observers may change the present result (the tentative conclusion). These three independent variables become input to decide wherein an event is fall to a given property or not. At this point, we also do not include a base value for a biased observer (Burghardt et al., 2012). The bias observer has a difference threshold in making a judgment, which leads by individualism (e.g. experience, knowledge and personal interest), internal influences (e.g. Ph.D. supervisor) and external influences (e.g. sponsor and government policy).

The USA-IND relations consist of a toggle event that is a property $\mathcal{P}_2$ (economic cooperation). The economy cooperation is stopped because of economy sanctions at a certain time frame and later it is revoked. The toggle event may enable and disable (or toggled multiple times) during the observation period. For an example, a diplomatic mission (or embassy) may open when a negotiation is still acceptable, but later it is closed due to relation crisis – how to evaluate this toggled event? Should an observer decide on enable or disable this property? or enable the property by divide it into a half of the property's value? – We have chosen to divide it by a half of the property's value. It is a trivial solution.

The news related to the international relations circled in a terrorism and global war on terrorism, conflicts in a middle east, nuclear and mass destruction weapon, head of state relation and sentiment, economy collaboration and sanctions, close border hostility and trespassing, diplomatic and espionage, and humanity and disaster relieve aid. We used these data as International Relation case study. Without the relation algebra, we can only use unsystematically way to identify relations between nations. For an example, refer to international relations between the USA and IND in the Case 3 section, one may directly interpret that both nations were in a hostile relation because of the President of the USA had said "… *decided to impose economy sanctions against India… Sanctions are mandatory under U.S. law when an undeclared nuclear state explodes a nuclear device.*". Through the relation algebra, we found that $R_{Mass}$ for $\mathcal{R}_{USA,IND}$ (0.135), which represent a fragile neutral relation. The relation algebra provides wider international relations assessments because it integrates multiple properties in its relation computation. The given tentative conclusions may change due to new evidence and new events that will be known by the observer in future. Furthermore, different observers may have different views regarding properties and events. One may use the Dempster-Shafer theory to evaluate (or combine) each observer's results. Later, one may further analyze the observer's results using Josang's subjective logic for an arithmetic and logic operations.

**Contribution and Future Work**

We have presented the relation algebra method for international nation relations. The proposed method has enabled relation computation, which is previously subjective and unquantifiable. The method seamlessly works together with the existing trust methodologies, which are Dempster-Shafer theory, Josang's subjective logic and Bayesian theory. We have also presented interesting case studies to demonstrate the practicality of the proposed method. We believe that our method can assist researchers studying in the field of international relations. Government officers working in foreign ministry or defense ministry may adopt our method as a quantitative methodology for international trust evaluations between foreign nations. Department of Defense may use our method to identify a nation that can be identified as trusted or neutral; or a hostile nation that can be anticipated as a terrorist nation or unsettled nation.

---

[7] We do not obtain or use any material that may lead to actions of a cyber-crime, terrorism, spying or any other illegal activities.

The most significant contribution of the proposed method that it will help to find the most trusted authorizing nation in Common Criteria. Trust is an important element to ensure Common Criteria's participant nations are able to recognize and consume Common Criteria's products. Choosing the most trusted authorizing nation for product evaluations will secure a value chain of the entire architecture of Comment Criteria. The relation algebra allows relation computations between nations that contribute to trust credentials in the Common Criteria's participant nations. The method allows one to test the Kallberg's hypothesis (Kallberg, 2012) regarding the *"long-term survival of CC requires abandoning the global approach and instead use established groupings of trust"*. To test the hypothesis, we need the relation computation method for international nation relations – that is using our method. Later, we need to evaluate trust relations between the Common Criteria's participant nations. Then, it will reveal that either the given hypothesis is valid or not.

Refer to the Literature Review section, we have done an exhaustive search for a method to perform the relation computation in the international nation relations to no avail. To the best of our knowledge, this is the first attempt in computer science in the area of information security research to model the relation algebra method for international nation relations. We also acknowledge that, the proposed method is not completed yet, and there are many research gaps and opportunities that are still available for the future research works.

**Conclusion**

In this work, we have modeled the relation algebra for international nation relations. The purpose of relation algebra method is to allow relation computations and trust modeling. Previously, there is no such a method to perform the relation computations for international relations which are subjective and unquantifiable. We have also presented the international nation relations between USA-GBR, USA-IRN and USA-IND as case studies to demonstrate the proposed method in a real-world scenario. We have met our research objectives by applying and verifying the relation algebra with the case studies. We plan to publish the relation algebra for Common Criteria's participant nations in the next publication.

**About the authors**

Mohd Anuar Mat Isa is a former researcher at Malaysia Institute of Microelectronic Systems (MIMOS) since 2008 until 2011. He graduated from Universiti Tenaga Nasional, Malaysia with the first class bachelor degree in Computer Science in 2008. Currently, he is a freelance researcher and involves in various research works. He is a member of the Malaysian Society for Cryptology Research (MSCR). He has published more than 20 research publications. He has also filed more than 10 patents and holding 2 software copyrights. His research focuses on Asymmetric Cryptography, Lightweight Cryptography for Embedded Devices, Embedded Firmware Security, Formal Methods, Trust and Mathematical Modeling, International Relations, and Trusted Computing.

Professor Ramlan Mahmod obtained his degree in Computer Science from Michigan State University, USA and his Master in Computer Science from Central Michigan University, USA. His Ph.D. is in Artificial Intelligence from Bradford University, United Kingdom. He has been a lecturer at Universiti Putra Malaysia 1985 and is currently the Dean of Computer Science Faculty, Universiti Putra Malaysia. He was seconded to MIMOS Berhad for two years from 2008 -2009 to help R&D in Trusted Computing and Information Security. He has published more than 75 journal papers and more than 110 articles in conference proceedings. He has filed 10 patents and holding more than 10 software copyrights. More than 25 Ph.D. and Master student graduated under his supervision and currently supervising and co-supervising more than 15 Ph.D. and Master student. His current research interest is Information Security especially in Cryptographic Algorithms, Steganography, Digital Forensics and Trusted Computing.

Assoc. Prof. Nur Izura Udzir is an academic staff at the Faculty of Computer Science and Information Technology, Universiti Putra Malaysia (UPM) since 1998. She received her Bachelor of Computer Science (1995) and Master of Science (1998) from UPM, and her Ph.D. in Computer Science from the University of York, UK (2006). She is a member of IEEE Computer Society and a Committee Member of Information Security Professionals Association of Malaysia (ISPA.my). Her areas of specialization are access control, secure operating systems, intrusion detection systems, coordination models and languages, and distributed systems. She is currently the Leader of the Information Security Group at the faculty.

Ali Dehghantanha obtained his Ph.D. in Computer Science with a specialization in Security in Computing from University Putra Malaysia (UPM). He now attached with school of Computing, Science and Engineering, University of Salford, Manchester, United Kingdom (U.K). His research interest in cyber forensics (malware analyzing, big-data investigation, SDN forensics, IoT investigation) cyber-crime (criminology and policy research), anti (online) money laundering and counter terrorism financing, and privacy issues in digital forensics.

Jamalul-lail Ab. Manan obtained his Ph.D. in Electrical Engineering from University of Strathclyde, UK. He joined MIMOS Berhad since 2006 until 2015 as Senior Director. His research works focus on Information Security and in particular in the areas of Trusted Computing Technologies, Privacy Enhancing Technologies and Data Protection Technologies.

Professor Audun Jøsang joined the University of Oslo in 2008. Prior to that he was Associate Professor at QUT, research leader of the Security Unit at DSTC in Brisbane, worked in the telecommunications industry for Alcatel in Belgium and for Telenor in Norway. He was also Associate Professor at the Norwegian University of Science and Technology (NTNU). He has a Master's in Information Security from Royal Holloway College, University of London, and a Ph.D. from NTNU in Norway. His research interests in Identity Management, Trust and Reputation Management, Network Security, Security Usability and Subjective logic.

**Acknowledgements**

The authors would like to acknowledge the Ministry of Education (MOE) Malaysia for providing research funding and Universiti Putra Malaysia (UPM) for supporting this research work.

**References**

Abdel-hafez, A., 2013. Trust, Privacy, and Security in Digital Business 8058. doi:10.1007/978-3-642-40343-9

Bansal, G., Mariam, F., 2015. Trust violation and repair : The information privacy perspective. Decis. Support Syst. 71, 62–77. doi:10.1016/j.dss.2015.01.009

BBC News, 2001. Afghanistan's Northern Alliance [WWW Document]. URL http://news.bbc.co.uk/2/hi/south_asia/1552994.stm (accessed 11.18.15).

Biskup, J., Hielscher, J., Wortmann, S., 2008. A Trust- and Property-based Access Control Model. Electron. Notes Theor. Comput. Sci. 197, 169–177. doi:10.1016/j.entcs.2007.12.026

Brewka, G., 1992. Nonmonotonic Reasoning: Logical Foundations of Commonsense, in: Cambridge University Press, ACM SIGART Bulletin. pp. 28–29.

Burghardt, G.M., Bartmess-LeVasseur, J.N., Browning, S.A., Morrison, K.E., Stec, C.L., Zachau, C.E., Freeberg, T.M., 2012. Perspectives - Minimizing Observer Bias in Behavioral Studies: A Review and Recommendations. Ethology 118, 511–517. doi:10.1111/j.1439-0310.2012.02040.x

Business Standard, 1998. Clinton Imposes Full Sanctions On India [WWW Document]. URL http://www.business-standard.com/article/specials/clinton-imposes-full-sanctions-on-india-198051401086_1.html (accessed 11.18.15).

Charter of the United Nations [WWW Document], 1945. URL http://www.un.org/en/documents/charter/ (accessed 6.11.15).

CNN, 1998. U.S. imposes sanctions on India [WWW Document]. URL http://edition.cnn.com/WORLD/asiapcf/9805/13/india.us/ (accessed 11.18.15).

Commerce, U.S.D. of, 2014. Trade in Goods with Iran [WWW Document]. URL https://www.census.gov/foreign-trade/balance/c5070.html (accessed 11.18.15).

Council on Foreign Relations, 2007. Agreement for Cooperation Between the Government of the United States of America and the Government of India Concerning Peaceful Uses of Nuclear Energy (123 Agreement) [WWW Document]. URL http://www.cfr.org/india/agreement-cooperation-between-government-united-states-america-government-india-concerning-peaceful-uses-nuclear-energy-123-agreement/p15459 (accessed 11.18.15).

Dan Balz and William Drozdiak, 1998. U.S. Responds With Penalties, Persuasion [WWW Document]. Washingt. Post Foreign Serv. URL http://www.washingtonpost.com/wp-srv/inatl/longterm/southasia/stories/penalties051498.htm (accessed 11.18.15).

Dempster, a P., 1967. Upper and lower probabilities induced by a multivariate mapping. Ann. Math. Stat. 38, 325–339. doi:10.1214/aoms/1177733256

Detlef F. Sprinz, Y.W., 2004. Case study methods: Design, use, and comparative advantages. Model. numbers, cases Methods Stud. Int. relations 19–55.

Eldad, A., 2013. The Israeli-Palestinian conflict is a war of religion, not territory [WWW Document]. URL http://www.haaretz.com/opinion/.premium-1.542196 (accessed 1.1.15).

Embassy of Switzerland -Foreign Interests Section [WWW Document], 2015. URL https://www.eda.admin.ch/countries/iran/en/home/representations/embassy-of-switzerland-foreign-interests-section.html (accessed 11.17.15).

Esther Pan, J.B., 2010. The U.S.-India Nuclear Deal [WWW Document]. Counc. Foreign Relations. URL http://www.cfr.org/india/us-india-nuclear-deal/p9663 (accessed 11.30.15).

Fang, H., Guo, G., Zhang, J., 2015. Multi-faceted trust and distrust prediction for recommender systems. Decis. Support Syst. 71, 37–47. doi:10.1016/j.dss.2015.01.005

Golbeck, J., 2006. Computing with trust: Definition, properties, and algorithms. 2006 Secur. Work. doi:10.1109/SECCOMW.2006.359579

Great Britain. Parliament. House of Commons. Foreign Affairs Committee, 2010. UK-US Relations : Sixth Report of the Foreign Affairs Committee of Session 2009-10.

Hallerstede, S., Jastram, M., Ladenberger, L., 2014. A method and tool for tracing requirements into specifications. Sci. Comput. Program. 82, 2–21. doi:10.1016/j.scico.2013.03.008

Hardin, R., 1992. The Street-level Epistemology of Trust.pdf. Polit. Soc. 21, 505–529.

Heitmeyer, C.L., 2009. On the Role of Formal Methods in Software Certification: An Experience Report. Electron. Notes Theor. Comput. Sci. 238, 3–9. doi:10.1016/j.entcs.2009.09.001

Interests Section of the Islamic Republic of Iran [WWW Document], 2015. URL http://www.daftar.org/Eng/default.asp?lang=eng (accessed 11.17.15).

J.Wheeler, N., 2012. Trust-Building in International Relations. Peace Prints South Asian J. Peacebuilding 4, 21.

Jelenc, D., Trček, D., 2014. Qualitative trust model with a configurable method to aggregate ordinal data. Auton. Agent. Multi. Agent. Syst. 28, 805–835. doi:10.1007/s10458-013-9239-8

Jiang, X., Jiang, F., Cai, X., Liu, H., 2015. How does trust affect alliance performance? The mediating role of resource sharing. Ind. Mark. Manag. doi:10.1016/j.indmarman.2015.02.011

Jøsang, A., 1997. Artificial Reasoning with Subjective Logic, in: Artificial Reasoning with Subjective Logic. pp. 1–17.

Jøsang, A., 2001. a Logic for Uncertain Probabilities. Int. J. Uncertainty, Fuzziness Knowledge-Based Syst. 09, 279–311. doi:10.1142/S0218488501000831

Jøsang, A., 2013. Subjective Logic, University of Oslo, Draft, 18 February 2013.

Jøsang, A., 2015. Subjective Logic, University of Oslo. doi:10.1016/j.artint.2007.04.006

Kallberg, J., 2012. Common Criteria meets Realpolitik - Trust, Alliances, and Potential Betrayal. Secur. Privacy, IEEE 10.

Kashani, M.R.A., 2015. Democracy, Media, and Soft Security Umbrella Pattern. J. Curr. Res. Sci. 3, 122–127.

Katzman, K., 2009. Iran: U. S. Concerns and Policy Responses.

Koster, A., Schorlemmer, M., Sabater-Mir, J., 2012. Engineering trust alignment: Theory, method and experimentation. Int. J. Hum. Comput. Stud. 70, 450–473. doi:10.1016/j.ijhcs.2012.02.007

Kydd, A.H.., 2005. Trust and Mistrust in International Relations. Princeton University Press.

Martinelli, F., Petrocchi, M., 2007. On Relating and Integrating Two Trust Management Frameworks. Electron. Notes Theor. Comput. Sci. 168, 191–205.

doi:10.1016/j.entcs.2006.12.005

McGillivray, F., Smith, A., 2000. Trust and cooperation through agent-specific punishments. Int. Organ. 54, 809–824.

Mcknight, D.H., Chervany, N.L., 1996. The Meanings of Trust. doi:10.1117/12.304574

Mcknight, D.H., Chervany, N.L., 2001. Trust and Distrust Definitions : One Bite at a Time. Trust Cyber-societies, Lect. Notes Comput. Sci. 2246, 27–54. doi:10.1007/3-540-45547-7_3

Menon, T., Sheldon, O.J., Galinsky, A.D., 2014. Barriers to Transforming Hostile Relations: Why Friendly Gestures Can Backfire. Negot. Confl. Manag. Res. 7, 17–37. doi:10.1111/ncmr.12023

Mir H. Sadat, Hughes, J.P., 2010. U.S.-Iran Engagement Through Afghanistan. Middle East Policy 17, 31–51.

Mohd Anuar Mat Isa, Jamalul-lail Ab Manan, Ramlan Mahmod, Habibah Hashim, Mar Yah Said, Nur Izura Udzir, Ali Dehghan Tanha, 2012a. Finest Authorizing Member of Common Criteria Certification, in: The International Conference on Cyber Security, CyberWarfare and Digital Forensic 2012 (CyberSec12). pp. 166–171.

Mohd Anuar Mat Isa, Jamalul-lail Ab Manan, Ramlan Mahmod, Habibah Hashim, Nur Izura Udzir, Ali Dehghan Tanha, Mar Yah Said, 2012b. Game Theory: Trust Model for Common Criteria Certifications & Evaluations. Int. J. Cyber-Security Digit. Forensics 1, 50–58.

Newton, K., 2001. Trust, social capital, civil society, and democracy. Int. Polit. Sci. Rev. 22, 201–214.

NSA, 2013. UKUSA [WWW Document]. URL https://www.nsa.gov/public_info/declass/ukusa.shtml (accessed 11.17.15).

Oelsner, A., 2007. Friendship, mutual trust and the evolution of regional peace in the international system. Crit. Rev. Int. Soc. Polit. Philos. 10.

Pollack, M.A., 2001. International Relations Theory and European Integration. J. Common Mark. Stud. 39, 245–64. doi:10.1111/1468-5965.00287

Porta, P.L., Scazzieri, R., 1997. Towards an economic theory of international civil society: Trust, trade and open government. Struct. Chang. Econ. Dyn. 8, 5–28. doi:10.1016/S0954-349X(96)00064-1

Public Statement on U.S. Policy Toward the Iran Nuclear Negotiations [WWW Document], 2015. URL http://www.washingtoninstitute.org/policy-analysis/view/public-statement-on-u.s.-policy-toward-the-iran-nuclear-negotiations (accessed 11.11.15).

Rod Nordland, T.N.Y.T., 2014. In Farewell Speech, Karzai Calls American Mission in Afghanistan a Betrayal [WWW Document]. URL http://www.nytimes.com/2014/09/24/world/asia/hamid-karzai-afghanistan.html?_r=0 (accessed 11.3.15).

Rohner, D., Thoenig, M., Zilibotti, F., 2012. War Signals : A Theory of Trade , Trust and Conflict. Rev. Econ. Stud. 80.

Schoorman, F., Mayer, R., Davis, J., 2007. An integrative model of organizational trust: Past, present, and future. Acad. M anagement Rev. 32, 344–354.

Shafer, G., 1976. A Mathematical Theory of Evidence. Princeton Univivserity Press.

Shafer, G., 1979. Allocations of Probability. Ann. Probab. 7.

Shafer, J., 2015. A Mathematical Theory of Evidence [WWW Document]. URL http://www.glennshafer.com/books/amte.html (accessed 1.1.15).

Sherman, W., 2013. U.S. Policy Toward Iran [WWW Document]. URL http://www.state.gov/p/us/rm/2013/202684.htm (accessed 11.17.15).

Svetlana, S., Tzukanova, D., 2016. Two Possible Ways of The Sociology Development as Two Versions of The Social Dimension. J. Curr. Res. Sci. 4, 47–51.

The "Grand Bargain" Fax: A Missed Oppurtunity [WWW Document], 2002. URL http://www.pbs.org/wgbh/pages/frontline/showdown/themes/grandbargain.html (accessed 11.17.15).

Trček, D., 2008. Managing Trust in Services Oriented Architectures, in: Proceedings of the 9th WSEAS International Conference on Applied Informatics and Communications. World Scientific and Engineering Academy and Society (WSEAS),. pp. 23–28.

Trček, D., 2009. A formal apparatus for modeling trust in computing environments. Math. Comput. Model. 49, 226–233. doi:10.1016/j.mcm.2008.05.005

Trček, D., 2011. Trust Management in the Pervasive Computing Era. IEEE Secur. Priv. 9.

US Department of the Treasury, 2014. Iran Sanctions [WWW Document]. URL http://www.treasury.gov/resource-center/sanctions/Programs/pages/iran.aspx (accessed 11.17.15).

Walt, S.M., 1998. International Relations: One World, Many Theories. Foreign Policy 29–46. doi:10.2307/1149275

Walter, F.E., Battiston, S., Schweitzer, F., 2008. A model of a trust-based recommendation system on a social network. Auton. Agent. Multi. Agent. Syst. 16.

White, H.C., 1992. Identity and Control: A Structural Theory of Social Action.

Wikipedia, 2015a. Iran–United_States_relations [WWW Document]. URL https://en.wikipedia.org/wiki/Iran%E2%80%93United_States_relations (accessed 11.17.15).

Wikipedia, 2015b. Northern Alliance [WWW Document]. URL https://en.wikipedia.org/wiki/Northern_Alliance (accessed 11.18.15).

Wobcke, W., 1995. Belief Revision, Conditional Logic and Nonmonotonic Reasoning. Notre Dame J. Form. Log. 36, 55–102. doi:10.1305/ndjfl/1040308829

Yan, Z., Holtmanns, S., 2007. Trust modeling and management: from social trust to digital trust. Comput. Secur. Priv. Polit. Curr. Issues, Challenges Solut. 27. doi:10.4018/978-1-59904-804-8.ch013

Yu, S., Beugelsdijk, S., de Haan, J., 2015. Trade, trust and the rule of law. Eur. J. Polit. Econ. 37, 102–115. doi:10.1016/j.ejpoleco.2014.11.003